\documentclass[12pt, draftclsnofoot, onecolumn]{IEEEtran}
\usepackage{filecontents}
\usepackage{amsthm}
\usepackage[fleqn]{amsmath}
\usepackage[table,xcdraw]{xcolor}
\newtheorem{theorem}{\textbf{\text{Theorem}}}
\newtheorem{corollary}{Corollary}
\newtheorem{lemma}{Lemma}
\usepackage{epsfig}
\usepackage{times}
\usepackage{verbatim}
\usepackage{enumerate}
\usepackage{multicol,afterpage,wrapfig}
\usepackage{cite}
\usepackage{graphicx}
\usepackage{algorithm}
\usepackage{algorithmic}
\usepackage{ifmtarg}
\usepackage{cite}
\usepackage{amssymb}
\usepackage{amsthm}
\usepackage[fleqn]{amsmath}
\usepackage{amsmath}
\usepackage{color}
\usepackage{bbm}
\usepackage{cite}
\usepackage{epstopdf}
\usepackage[table,xcdraw]{xcolor}
\usepackage{flushend}
\usepackage{caption}
\usepackage{subcaption}
\newtheorem{definition}{\textbf{\text{Definition}}}

\makeatletter
\newcommand*{\rom}[1]{\expandafter\@slowromancap\romannumeral #1@}
\makeatother

\begin{document}
	\title{A Unified Asymptotic Analysis of Area Spectral Efficiency in Ultradense Cellular Networks }
	\author{
		\IEEEauthorblockN{\large Ahmad AlAmmouri, Jeffrey G. Andrews, and Fran\c cois Baccelli}
		\thanks{The authors are with the Wireless Networking and Communications Group (WNCG), The University of Texas at Austin, Austin, TX 78701 USA. (Email: \{alammouri@utexas.edu, jandrews@ece.utexas.edu, francois.baccelli@austin.utexas.edu\}). Part of this work will be presented at IEEE International Symposium on Information Theory (ISIT), 2018 \cite{Asymptotic_AlAmmouri18}. Last revised \today.}}
	
	\maketitle
	
	\begin{abstract}
This paper studies the asymptotic properties of { average} area spectral efficiency (ASE) of a downlink
cellular network in the limit of very dense base station (BS) and user densities. 
This asymptotic analysis relies on three assumptions: (1) 
interference is treated as noise; (2) the BS locations are drawn from a Poisson point process;
(3) the path loss function is bounded above satisfying mild regularity conditions.
We consider three possible definitions of the { average} ASE, all of which give units of bits per second {per unit bandwidth} per unit area. 
When there is no constraint on the minimum operational {signal-to-interference-plus-noise ratio} (SINR)
and instantaneous full {channel state information} (CSI) is available at the transmitter,
the { average} ASE is proven to saturate to a constant, which we derive in a closed form. 
For the other two ASE definitions, wherein either a minimum SINR is enforced or CSI 
is not available, the { average} ASE is instead shown to collapse to zero at high BS density. 
We provide several familiar case studies for the class of considered path loss models, 
and demonstrate that our results cover most previous models and results on
ultradense networks as special cases.  
	\end{abstract}
	
	\section{Introduction}
The reuse of spectrum across space, called frequency or spatial reuse, is historically
the most important attribute of terrestrial wireless networks. Installing new cellular or
wireless LAN infrastructure is expected to generate spatial reuse, whereby the time and frequency
resources can be reused across smaller spatial scales. A major question within this
context is whether this will indefinitely bring higher overall data throughputs in the network. 
The area spectral efficiency (ASE), defined as the sum of the maximum data rate per unit area per unit bandwidth
\cite{Area_Alouini99}, has historically increased about linearly with the number of base stations (BSs):
doubling the number of BSs about doubles the sum throughput the network supports in a given area. 
This has often been referred to as ``cell splitting gain'' or ``densification gain'' and has been
the most significant driver in increased wireless data rates over the last few decades
\cite{ChaAnd08,CoopersLaw}, and is expected to continue to drive gains for the foreseeable
future as well \cite{Qualcomm15,What_Andrews14}. The aim of the present paper is to clarify what 
can in fact be expected in terms of ASE under densification. This is done by the definition
of a general model allowing one to study the asymptotic behavior of ASE under broad conditions.

	\subsection{Related Work}

The study of dense wireless network capacity has a rich history, in particular for
the case of ad hoc multihop networks.  The seminal result of Gupta and Kumar \cite{GupKum00} 
showed that the transport capacity of an infrastructure-less network increases with the number
of nodes $n$ roughly as $\sqrt{n}$, assuming a given node wishes to transmit with a
randomly selected other node.  Many subsequent results extended this ``scaling law''
approach to a wide variety of models and communication techniques, well summarized in \cite{XueKum06}.
A main intuition from these results is that the best network-wide throughput scaling
is achieved by short-range local communication coupled with multi-hopping towards the
intended receiver, a form of spatial reuse. A key subsequent paper showed that this
square-root scaling is a fundamental property of the electromagnetic propagation \cite{Fra07},
thus attempts to improve the scaling law through exotic communication strategies would be futile.  

To obtain mathematically tractable results on wireless network performance without resorting to
scaling laws, stochastic geometry \cite{Stochastic_Baccelli10,Stochastic_Baccelli10_2,HaenggiBook,Modeling_ElSawy16}
has emerged as a key toolset, in particular for deriving the {signal-to-interference-plus-noise ratio} (SINR)  distribution and related metrics,
as achieved for ad hoc networks in \cite{An_Baccelli06,WebYan05}.  
One of the key conclusions of early stochastic geometry results for cellular networks
such as\cite{A_Andrews11,Modeling_Dhillon12} was that the SINR distribution is 
insensitive to the BS density, once the network is interference-limited.  
This indicated that the {  average} ASE should increase linearly with the BS density: an encouraging result.  
Both \cite{A_Andrews11,Modeling_Dhillon12}, as well as most prior and subsequent
literature including the aforementioned scaling laws results, use a standard power-law path
loss function of the form $L(r) = c_0 r^{-\eta}$, where $r$ is the distance,
$c_0 = L(1)$ is a constant, and $\eta > 2$ is the path loss exponent.  

However, other more recent works have shown that when adopting more 
realistic bounded path loss models, different and much more pessimistic trends
can be derived or observed for {  average} ASE under densification
\cite{Downlink_Atzeni17,SINR_AlAmmouri17,Downlink_Zhang15,Performance_Ding16,The_Renzo16,Spectral_Lee16,Performance_Nguyen17,Stochastic_Filo17}.  
The current situation can be summarized by saying that one finds in the literature
the whole spectrum of asymptotic behaviors for ASE ranging from linear growth to collapse.
{  The reasons behind having different asymptotic behaviors are mainly due to the used path loss model and the used metric to measure the average ASE. Different assumptions regarding the channel state information (CSI) availability and minimum operational SINR have been used in the literature, and changing these assumptions might lead to a totally different scaling law. Hence, we dedicate Section \ref{Sec:ASE} to rigorously defining the average ASE under different assumptions for the CSI availability and the minimum operational SINR required for a successful  transmission.

Regarding the path loss model,  it is well understood that the pole at the origin of the standard power-law model
creates a scale-free cellular SIR which is the basis for the linear ASE growth alluded to above. 
It is also well understood that this standard power-law path loss model has important shortcomings 
in the context of the modeling of dense networks: first, it does not accurately model received powers
for short distances, and second, it is intractable for $\eta = 2$. 
The more realistic bounded path loss models alluded to above, which are the main focus of the present paper,
are hence much more natural within the context of ultra-densification. To avoid distracting the readers by listing the  different path loss models used in the literature and the different definitions of the ASE, we discuss the results available in the literature and connect them with the results in this work in Section \ref{Sec:Examples}.}

	\subsection{Contributions}
The main contribution of the paper is a general answer to the question of how {  average} ASE scales in a cellular network
with very large density. We start by formally defining the {  average} ASE using the Shannon rate,
with the important but common caveat that interference is treated as noise. 
We further provide two alternative definitions that add practical constraints
regarding the minimum operational SINR and the availability of the channel state
information at the transmitters.
We then define a broad class of bounded path loss models
that would appear to capture all currently used and physically viable propagation models. 
This broad class, which is assumed throughout the paper, is defined by three simple mathematical
properties, the most important being boundedness.
Then, under a Poisson point process (PPP) assumption regarding the spatial
distribution of the BSs and a general small-scale fading model,
we derive the following properties on the asymptotics of the different definitions
of the ASE when the BS density grows large. 
	
Assuming that there is no constraint on the minimum operational SINR, and that
the transmitter can send at the Shannon rate -- which implies that
perfect instantaneous CSI is available at the transmitter --
we prove that the {  average} ASE \emph{saturates to a constant},  which is $L_0/2\pi \ln(2) \gamma$,
where $L_0$ and $\gamma$ are constants determined by the path loss function $L(r)$,
namely $L_0<\infty$ is the path loss at $r=0$, and $\gamma = \int\limits_{0}^{\infty} r L(r)dr$.  
However, under the same set of assumptions, when either a minimum operational SINR is required,
or the CSI is not fully available, then the {  average} ASE tends to zero when the BS density tends to infinity.

The rest of the paper is organized as follows. In Section \ref{Sec:SM}, we present the system model.
In Section \ref{Sec:ASE}, we present the different definitions of the ASE and discuss the 
intuition behind each one of them. Section \ref{Sec:PerAnalysis} is the main technical section of the paper,
where we derive expressions for the {  average} ASE. Case studies with discussions are presented
in Section \ref{Sec:Examples}, before the conclusion and future work in Section \ref{Sec:Conc}.
	
	\textit{\textbf{Notation}}: $\mathbbm{1}{\{\cdot\}}$ is the indicator function which takes the value $1$ if the statement $\{\cdot\}$ is true and takes the value $0$ otherwise, $\mathbb{R}$ is the set of real numbers, $\mathbb{Z}$ is the set of integers, $\mathbb{R}_{+}$ is the set of non-negative real numbers, and $\mathbb{R}^{*}_{+}$ is the set of strictly positive real numbers.
	
\section{System Model}\label{Sec:SM}
We consider a downlink cellular network where the BSs are spatially distributed as a homogeneous PPP $\Phi$ with intensity $\lambda$. The users are spatially distributed according to an independent stationary point process, with intensity $\lambda_u \gg \lambda$ such that each BS has at least one user to serve { with probability one} . In other words, we assume that all BSs are continually transmitting. The small scale fading $h$ is assumed to be distance independent with a unit mean but with arbitrary distribution. All channel fading variables are assumed to be independent and identically distributed (i.i.d.).  Closest BS association is assumed and all BSs and users are equipped with a single omni-directional antenna. 

{  Since we are interested in the scaling laws with $\lambda$, we assume that $\lambda_u$ scales linearly with $\lambda$ such that all BSs have at least one user to serve with probability one. Moreover, we assume that each BS schedules its users on orthogonal resource blocks such that one user is associated with every BS in a given resource block. Hence, the users do not suffer from intra-cell interference, but they are still affected by the interference from other cells. Based on this, the intensity of active users in a given resource block is equal to $\lambda$.}

\textbf{Propagation Model.} The large scale channel gain is captured by the path loss function $L(r)$.  We focus on a broad class of path loss models that are physically reasonable. First, we assume that $L(0)=L_0$ is the average transmit power directly at the antenna, hence $L_0$ is a finite constant. Moreover, we assume that $L(r)\leq L_0 \ \forall \ r \in  \mathbb{R_{+}}$, which means that at any distance, the average received power cannot be higher than the average transmit power present at the antenna itself. 
Note that assuming $L(r)\leq L_0 \ \forall \ r \in \mathbb{R_{+}}$ is much weaker than requiring  $L(r)$ to be decreasing with the distance. Hence, this assumption is quite general and includes cases where the path loss is not deterministic with the distance, such as the common situation where depending on a random link state (e.g. LoS or NLoS), different path loss attenuation functions are activated \cite{Analysis_Bai14,RanRapErk14,Performance_Ding16}.

Finally, we require that the total average received power $P_{\rm avg}$ at the origin (or any arbitrary point in the network) in the network to be finite. This requirement can be mathematically distilled as
\begin{align} \label{Eq:Pavg}
P_{\rm avg}=\mathbb{E}\left[ \sum\limits_{r_i\in \Phi} h_i L(r_i)\right]=\mathbb{E}\left[ \sum\limits_{r_i\in \Phi}  L(r_i)\right]=2 \pi \lambda \int\limits_{0}^{\infty} r L(r)dr=2 \pi \lambda \gamma,
\end{align}
where $\gamma \triangleq \int\limits_{0}^{\infty} r L(r)dr$. To obtain this, we used the fact that the small scale fading has unit mean, and then Campbell's theorem \cite{Stochastic_Baccelli10} to get the last equality. Hence, we assume that $\gamma$ is finite. {   In other words, the path loss function has to be integrable over $\mathbb{R}^{2}$. Note that given the first (boundedness) and the second (non-increasing) requirements, the third puts a restriction on how fast $L(\cdot)$ drops to zero. In fact, if the path loss model is bounded but the integral in \eqref{Eq:Pavg} is not finite, then the received interference at any point in the network is infinite almost surely and not only on average \cite[Theorem 4.6]{HaenggiBook}. For example, $L(r)=(1+r)^{-\eta}$ satisfies the first two requirements for all $\eta \geq 0$, but $\eta$ has to be larger than $2$ in order for the last condition to be satisfied. But, it is well-known that for the case of $\eta \leq2$, the interference at any point in the network is infinite almost surely \cite[Section 5.1]{HaenggiBook}. However, this is only true for bounded path loss functions. If the path loss is not bounded, then it is possible to have a finite interference almost everywhere in the network, but on average the received interference has an infinite mean as in the case of the power-law path loss $L(r)=r^{-\eta}$ with $\eta>2$. The scaling laws in this case are totally different from the laws derived in this work\cite{A_Andrews11,Scaling_Baccelli15}.}

We formalize the preceding discussion with a definition of physically feasible path loss models.
\begin{definition}\label{Def:PLM}(Physically feasible path loss)
	Physically feasible path loss models are the family of path loss functions $L(r)$ with the following properties 
	\begin{enumerate}
		\item $L_0=L(0) \in \mathbb{R}_{+}$,
		\item $L(r)\leq L_0 \ \forall r\in \mathbb{R}_{+}$,
		\item $\gamma=\int\limits_{0}^{\infty} r L(r)dr \in \mathbb{R^{*}_{+}}$,
	\end{enumerate}
	\end{definition}
  
To summarize concisely the rationale for these three conditions:
1) ensures that the transmit power is finite, 2) ensures that on average we cannot receive more power than was transmitted, and 3) states that the average received power at any point from all BSs in the network is finite. Hence, we conclude that it is impossible to construct an empirically verifiable path loss model which violates any of these three conditions in Definition \ref{Def:PLM}.  Unsurprisingly, the path loss models adopted in 3GPP standards, both for the conventional Sub6 GHz \cite{3GPP2010} as well as in the mmWave bands \cite{3GPP2017}, along with a great many other common path loss models, satisfy the aforementioned requirements as we will show in Sect. \ref{Sec:Examples}. Throughout this work, all results are for this class of physically feasible path loss models, unless otherwise stated. Moreover, we allow for any general small-scale fading model as long as it has finite mean, for simplicity we assume unit mean\footnote{The analysis can be easily extended to the non-unit mean fading distributions by taking the mean as a common factor in the SINR expression in \eqref{eq:SINRDef}. Hence, it is equivalent to considering unit mean distributions with a scaled noise power.}. {  Note that if $L(\cdot)$ is not deterministic with the distance, then $\gamma$ in Definition \ref{Def:PLM} is equal to $\mathbb{E}[\int_{0}^{\infty} r L(r)dr ]$, where the expectation is with respect to the randomness in $L(\cdot)$.}

\textbf{SINR.} We derive the performance of a user located at the origin. According to Slivnyak's theorem \cite{Stochastic_Baccelli10_2}, there is no loss of generality in this assumption, and the evaluated performance represents the average performance for all the users in the network. Based on our system model, the SINR for the typical user is
\begin{align}
{\rm SINR}=\frac{h_0 L(r_0)}{\sum\limits_{r_i\in \Phi \setminus B(0,r_0)} h_i L(r_i)+N_0 },\label{eq:SINRDef}
\end{align}
where $N_0$ is the noise power, $h_0$ ($h_i$) and $r_0$ ($r_i$) are the channel small scale fading power and the distance between the tagged user and the serving ($i^{\text{th}}$ interfering) BS, respectively. Hence, we focus on the case where the network interference is treated as noise which is the baseline practice even in 5G systems. The same assumption was made in \cite{Area_Alouini99}, where the ASE was proposed.

{  However, treating interference as noise inherently assumes that the interference does not have any structure, while it actually has a structure that can be exploited to decode the desired messages\cite{Gaussian_Etkin08}.  From a decoding perspective, treating interference as noise is only known to be optimal for the sum rate in the case of a weak interference channel (the desired signal power is much larger than the interfering signals). For the other cases, joint decoding or successive interference cancellation (SIC) are known to achieve better rates than treating interference as noise \cite{Network_ElGamal11}. If we add possible cooperation between the BSs, then multiple BSs could cooperate to serve an edge user in addition to their own users. However, such a cooperation complicates the analysis and might lead to losing tractability. An attempt to solve a similar problem can be found in \cite{A_Baccelli15} by just considering coordination between two BSs, but the resulting expressions are complicated and it is hard to identify the asymptotics from them. Hence, we continue with treating interference as noise and we postpone considering more advanced decoding or cooperative techniques, as in \cite{A_Baccelli15,An_Huang13,Transmission_Weber07,Interference_Baccelli11,The_Zhang14}, for  future work.} Before delving into the analysis, we formally introduce different definitions of the ASE in the next section and discuss the intuition and the physical meaning behind them.

\section{Throughput performance metrics}\label{Sec:ASE}
The {  average} ASE has been widely used to study different trade-offs in cellular networks and as a measure of the total network throughput. The definition in \cite{Area_Alouini99} is based on the sum of the {  spectral efficiency (Shannon limit)} of the users per unit area:
\begin{align}\label{eq:ASE_Basic}
\mathbb{E} \left[ \mathcal{E}(N)\right]= \frac{1}{|\mathcal{A}|} \sum_{k=1}^{N} \mathbb{E} \left[\log_2(1+{\rm SINR}_k)\right],
\end{align}
where $\mathcal{A}$ is the considered area, $N$ is the number of users within $\mathcal{A}$, and $\log_2(1+{\rm SINR}_k)$ is the spectral efficiency in bps/Hz. Then by averaging over different fading realizations, we get the expression in \eqref{eq:ASE_Basic}. Note that the used Shannon {  limit} formula $\log_2(1+{\rm SINR}_k)$ assumes Gaussian interference, which is not true in our case since the interference is typically dominated by near BSs. Hence, the central limit theorem does not apply. However, Gaussian interference is the worst-case scenario for capacity calculations {  under the assumption of point-to-point coding and decoding (no BS cooperation), as mentioned earlier} \cite{Area_Alouini99,Information_Shamai97}.

 Moreover, there is randomness in the BSs/users locations and in the small-scale fading in our case. But since the SINR distribution seen by the typical user at the origin represents the stationary distribution of all users in the network, and since the channels are assumed to be i.i.d, then after averaging over different network and fading realizations, the definition \eqref{eq:ASE_Basic} simplifies to the following:
\begin{align}
	\mathbb{E} \left[\mathcal{E}(\lambda)\right]&=\lambda\mathbb{E}^{0} \left[ \log_2 (1+{\rm SINR})\right] ,\label{Eq:ASE_Def1_0}
\end{align}
where $\mathbb{E}^{0} [\cdot]$ is the Palm expectation with respect to the users' point process \cite{Stochastic_Baccelli10_2}. However, since we are focusing on a user located at the origin, the Palm expectation is reduced to the (stationary) expectation $\mathbb{E} [\cdot]$ due to Slivnyak's theorem \cite{Stochastic_Baccelli10_2}.  {  Note that $\lambda$ appears in \eqref{Eq:ASE_Def1_0} because the density of active users using a given resource block is the same as the density of BSs as discussed in Section \ref{Sec:SM}.} Hence, we adopt the following definition of the {  average} ASE.

\begin{definition}
	(ASE) The average ASE is defined as:
	\begin{align}
	\mathbb{E} \left[\mathcal{E}(\lambda)\right]&=\lambda\mathbb{E} \left[ \log_2 (1+{\rm SINR})\right] .\label{Eq:ASE_Def1}
	\end{align}
\end{definition}

The {  average} ASE definition in this form has been used in \cite{Spectral_Lee16,Area_Chen17,Asymptotic_Alam17,The_Renzo16,SINR_AlAmmouri17}. However, it assumes that the system can work with any arbitrarily small SINR, which may not be feasible in practice. Hence, a second definition was used in \cite{Performance_Ding16,Performance_Ding17}, which adds a constraint on the minimum operational SINR $\theta_0$ as in the next definition.

\begin{definition}
	(Constrained ASE) The average constrained ASE is defined as:
\begin{align}
\mathbb{E} \left[\tilde{\mathcal{E}}(\lambda,\theta_0)\right]&=\lambda\mathbb{E} \left[ \log_2 (1+{\rm SINR})\mathbbm{1} \left\{{\rm SINR}\geq\theta_0 \right\} \right] .\label{Eq:ASE_Def2} 
\end{align}
\end{definition}

 In addition, a third definition called the {  average} potential throughput, was  proposed in \cite{Downlink_Zhang15}, and then used in \cite{Downlink_Atzeni17,Success_Li16,The_Renzo16,SINR_AlAmmouri17,Modeling_Afshang16}, to study the {  average} ASE in cellular networks as the following:
	
	\begin{definition}
		(Potential throughput) The average potential throughput is defined as:
		\begin{align}
\mathbb{E} \left[\mathcal{R}(\lambda,\theta_0)\right]&=\lambda \log_2 (1+\theta_0) \mathbb{P} \left\{{\rm SINR}\geq\theta_0 \right\}  .\label{Eq:ASE_Def3} 
\end{align}
	\end{definition}

	 The {  average} potential throughput captures the case where the channel state information is not available at the transmitter, hence it transmits with a constant rate and then the messages are only decodable at the receiver if the SINR is larger than some threshold $\theta_0$. Which means that high SINR values are not exploited, and if the SINR is small, then the link is in complete outage. The three different definitions are related as follows:
	 \begin{align}
	 \mathbb{E} \left[\mathcal{R}(\lambda,\theta_0)\right]&=\mathbb{E} \left[\lambda \log_2 (1+\theta_0) \mathbbm{1} \left\{{\rm SINR}\geq\theta_0 \right\} \right]\notag\\
	 &\leq \mathbb{E} \left[\lambda \log_2 (1+{\rm SINR}) \mathbbm{1} \left\{{\rm SINR}\geq\theta_0 \right\} \right]=	\mathbb{E} \left[\tilde{\mathcal{E}}(\lambda,\theta_0)\right]\notag\\
	 &\leq \mathbb{E} \left[\lambda \log_2 (1+{\rm SINR}) \right]=	\mathbb{E} \left[\mathcal{E}(\lambda)\right]\notag.
	 \end{align}
	 
	 Hence $\mathbb{E} \left[\mathcal{R}(\lambda,\theta_0)\right] \leq   	\mathbb{E} \left[\tilde{\mathcal{E}}(\lambda,\theta_0)\right]\leq \mathbb{E} \left[\mathcal{E}(\lambda)\right]$ for all $\theta_0 \geq0$. In all cases, the BSs are assumed to always have messages to transmit to their users.

It was shown in \cite{SINR_AlAmmouri17} that adopting different definitions can lead to different insights on the network performance. The aim of this work is to study all of them in a unified framework. In the next section, we study the asymptotic value of the {  average} ASE under fairly general assumptions.

 \section{Performance Analysis}\label{Sec:PerAnalysis}
 In this section, we analyze the asymptotic behavior of cellular networks in terms of the {  average} ASE when the BS density is high. 
 Note that due to the boundedness of $L(\cdot)$, the numerator of the SINR defined in \eqref{eq:SINRDef} is bounded on average. However, the denominator can get arbitrary large by increasing the BS density. Hence, the SINR approaches zero for high BS density. For completeness, we prove this in the next lemma.

 \begin{lemma}
 	When $\lambda \rightarrow \infty$, the SINR as defined in \eqref{eq:SINRDef} tends to zero a.s.
   	\begin{proof}
   			Refer to Appendix A.
   		\end{proof}
 \end{lemma}
Next, we study the {  average} ASE as defined in \eqref{Eq:ASE_Def1} since it is the more general metric and then we move to the {  average} constrained ASE \eqref{Eq:ASE_Def2} and the {  average} potential throughput \eqref{Eq:ASE_Def3}. We start with the following lemma which provides a lower bound on the ASE.

 \begin{lemma}
 	The asymptotic average ASE is lower bounded by
 	\begin{align}\label{eq:ASELB}
 	\lim_{\lambda \rightarrow \infty}	\mathbb{E}\left[\mathcal{E}(\lambda) \right]\geq \frac{L_0}{2 \pi \ln(2) \gamma}.
 	\end{align}
 	\begin{proof}
 		Let $\lambda=k \lambda_0$, where $k \in \mathbb{Z}_{+}$ and $\lambda_0 \in \mathbb{R}^{*}_{+}$. We are interested in $\lim\limits_{k\rightarrow \infty} \mathcal{E}(k\lambda_0)$ which is found by the following
 		\begin{align}
 		\lim\limits_{k\rightarrow \infty} \mathcal{E}(k\lambda_0)&=\lim\limits_{k\rightarrow \infty} \lambda_0 k \log_2(1+{\rm SINR}(k\lambda_0))\notag\\
 		&=\lim\limits_{k\rightarrow \infty}\frac{\lambda_0}{\ln(2)}  k {\rm SINR}(k\lambda_0)\label{eq:Le1_1}\\
 		&=\frac{h_0 L_0}{\ln(2) 2 \pi \gamma}\label{eq:Le1_2}\\
 		\mathbb{E}\left[ \lim\limits_{k\rightarrow \infty} \mathcal{E}(k\lambda_0) \right]&=\frac{L_0}{\ln(2) 2 \pi \gamma}\label{eq:Le1_3},
 		\end{align}
 		where, \eqref{eq:Le1_1} follows because the SINR approaches zero a.s when $k \rightarrow \infty$ as we proved in Lemma 1, \eqref{eq:Le1_2}  follows from \eqref{eq:Le0_6}, and \eqref{eq:Le1_3}  follows by taking the expectation of \eqref{eq:Le1_2}  w.r.t $h_0$ which has a unit mean. Then by using Fatou's lemma \cite{Real_Royden88}, the following is true
 		\begin{align}
 		\lim\limits_{k\rightarrow \infty}\mathbb{E}\left[ \mathcal{E}(k\lambda_0) \right]\geq \mathbb{E}\left[ \lim\limits_{k\rightarrow \infty} \mathcal{E}(k\lambda_0) \right]=\frac{L_0}{\ln(2) 2 \pi \gamma}. \notag
 		\end{align}
 		
 		Note that the result is independent of $\lambda_0$, hence we can conclude that $\lim\limits_{\lambda \rightarrow \infty}\mathbb{E}\left[ \mathcal{E}(\lambda) \right]\geq\frac{L_0}{\ln(2) 2 \pi \gamma}$.
 	\end{proof}
 \end{lemma}
The last lemma shows that the average ASE is lower bounded by a constant and does not drop to zero. However, it is more interesting to show that it holds with equality. Then we will have exact characterization of the limit and prove the existence of a densification plateau \cite{Are_Andrews16}. In the following theorem, we provide a sufficient condition that can be tested without dealing with the limit of the ASE.
\begin{theorem}
	Let $\mathcal{I}=\sum_{r_i \in \Psi\setminus B(0,r_0)} L(r_i) h_i$, where $\Psi$ is a PPP with intensity $\lambda_0$. If the second negative moment of $\mathcal{I}$ is finite for all $\lambda_0\geq\lambda_c$, where  $\lambda_c\in \mathbb{R_{+}}$ is a constant, then
	\begin{align}
	\lim\limits_{\lambda \rightarrow \infty}\mathbb{E}\left[ \mathcal{E}(\lambda) \right]= \frac{L_0}{2 \pi \ln(2) \gamma}\label{eq:Th1_1}.
	\end{align}
	\begin{proof}
	{  The sketch of the proof is as follows: In the last step of the proof of Lemma 2, we used Fatou's lemma to show that $\lim\limits_{k\rightarrow \infty}\mathbb{E}\left[ \mathcal{E}(k\lambda_0) \right]\geq \mathbb{E}\left[ \lim\limits_{k\rightarrow \infty} \mathcal{E}(k\lambda_0) \right]$. However, it holds with equality if $ \mathcal{E}(k\lambda_0) $ is uniformly integrable for all $k$. Hence, in this proof, we show that, if the second negative moment measure of $\mathcal{I}$ is finite, then the uniform integrability condition is satisfied. For more details, refer to Appendix B}.
	\end{proof}
\end{theorem}

Theorem 1 shows that the {  average} ASE converges to a finite constant, which practically means that we cannot keep harvesting performance gains by densifying the network; at some point the performance will saturate. Note that Theorem 1 holds as long as the second negative moment of $\mathcal{I}$ as defined in the theorem is finite, which is a function of the path loss function as well as the fading distribution. In the next corollary, we provide two sufficient conditions under which the condition in Theorem 1 is satisfied.
\begin{corollary}
	If any of the following is satisfied:
	\begin{enumerate}
		\item $\int\limits_{0}^{\infty}\int\limits_{0}^{\infty} 2 \pi \lambda_0 r t \exp \left( -\pi \lambda_0 r^2 -2 \pi \lambda_0  \int_{r}^{\infty}x \mathbb{E}_h \left[1-  e^{-h t L(x)} \right] dx  \right)drdt$ is finite $\forall \lambda_0 \geq \lambda_c$,
		\item $\int\limits_{0}^{\infty} \int\limits_{0}^{\infty}2 \pi \lambda_0 r \exp \left(- \pi \lambda_0 r^2- 2 \pi \lambda_0 \int\limits_{r}^{\infty}x   \mathbb{P}_h\left(L(x) h \geq\frac{1}{\sqrt{t}} \right) dx \right) drdt$ is finite $\forall \lambda_0 \geq \lambda_c$,
	\end{enumerate}
	where $ \lambda_c\in \mathbb{R}_{+}$ is a constant, then the condition in Theorem 1 is satisfied.
	\begin{proof} 
	{  The proof follows by showing that if any of these conditions is satisfied, then $\mathcal{I}$, as defined in Theorem 1, has a finite second negative moment. For more details, refer to Appendix C.}
	\end{proof}
\end{corollary}
 { The conditions in the previous Corollary are listed for the case where $L(\cdot)$ is deterministic with the distance. Otherwise, the expectation in the first condition and the probability in the second condition, become with respect to $L$ in addition to $h$.} Although these conditions are general, evaluating these integrals for each path loss function and each fading distribution is cumbersome. Hence, in pursuit of simpler sufficient conditions, we present the following corollary for the Rayleigh fading case.
\begin{corollary}
	For Rayleigh fading channels (i.e. $h_i \sim \exp(1)$), if $\exists \  r_0 \in \mathbb{R}_{+}$ and $\zeta \in \mathbb{R}_{+}^{*}$ such that:
	\begin{enumerate}
		\item $\frac{rL(r)}{-L^{'}(r)} \geq \zeta ,\  \forall r \in [r_0,\infty)$,
		\item $\int\limits_{r_0}^{\infty} \frac{r}{L(r)^2} e^{- \pi \lambda_0 r^2}dr$ is finite for all $\lambda_0>\lambda_c\in \mathbb{R_{+}}$,
	\end{enumerate}
assuming that $L(r)$ is decreasing and differentiable $ \forall r\in[r_0,\infty)$, {  and deterministic with the distance,} then the first sufficient condition in Corollary 1 is satisfied. 
\begin{proof}
{  The idea in this proof is to show that if these conditions are satisfied, then the first condition in Corollary 1 is satisfied. For more details, refer to Appendix D.}
\end{proof}
\end{corollary}

The last corollary presents simple conditions that can be easily checked for any path loss function. Interestingly, these conditions cover most of the bounded path loss models reported in the literature as we will show in the next section. In the next corollaries, we show that these conditions are also sufficient for the no fading case as well as for any fading distribution as long as it has unit mean. 

\begin{corollary}
The sufficient conditions in Corollary 2 also hold for the no fading case.
	\begin{proof}
	For the no fading case, the first condition in Corollary 2 reduces to
	\begin{align}
	\mathbb{E}\left[ \frac{1}{\mathcal{I}_i^2}\right]&=\int\limits_{0}^{\infty}\int\limits_{0}^{\infty} 2 \pi \lambda_0 r t \exp \left( -\pi \lambda_0 r^2 -2 \pi \lambda_0  \int_{r}^{\infty}x \left(1-  e^{- t L(x)} \right)dx  \right)drdt\notag\\
	&\leq \int\limits_{0}^{\infty}\int\limits_{0}^{\infty} 2 \pi \lambda_0 r t \exp \left( -\pi \lambda_0 r^2 -2 \pi \lambda_0  \int_{r}^{\infty}x \frac{t L(x)}{1+ t L(x)}dx  \right)drdt,\label{Eq:Le4_2}
	\end{align}
	where \eqref{Eq:Le4_2} follows since $1-e^{-t}\geq \frac{t}{1+t}, \ \forall t\geq0$. Then the proof follows as in the proof of Corollary 2.
	\end{proof}
\end{corollary}  

\begin{corollary}
	The sufficient conditions in Corollary 2 hold for any fading model such that $\mathbb{E} \left[ h\right]=1$.
	\begin{proof} 
		Refer to Appendix E.
	\end{proof}
\end{corollary}  

Hence, the asymptotic behavior for the {  average} ASE is actually agnostic to the fading distribution. In other words, fading only quantitatively affect the performance for low to moderate densities. The conditions in Corollary 2 are for the case of a deterministic path loss function with the distance. However, if LoS/NLoS links are considered, then the path loss is no longer deterministic and it depends on the state of the link. This case is important when the network is operating in high frequency bands (e.g. mmWave) \cite{Modeling_Andews16,3GPP2017}. For this specific case, we provide the following corollary.


\begin{corollary}
	If $\exists \  r_0 \in {R}_{+}$,  $\zeta \in \mathbb{R}_{+}^{*}$ and a differentiable decreasing function $\tilde{L}: [r_0,\infty)\rightarrow \mathbb{R}_{+}$ such that:
	\begin{enumerate}
		\item $\tilde{L}(r)\leq L(r) \ \forall r \in [r_0,\infty)]$.
		\item $\frac{r\tilde{L}(r)}{-\tilde{L}^{'}(r)} \geq \zeta ,\  \forall r \in [r_0,\infty)$.
		\item $\int\limits_{r_0}^{\infty} \frac{r}{\tilde{L}(r)^2} e^{- \pi \lambda_0 r^2}dr$ is finite for all $\lambda_0>\lambda_c\in \mathbb{R_{+}}$.
	\end{enumerate}

	Then the first sufficient condition in Corollary 1 is satisfied.
	\begin{proof} The proof is straightforward from the first sufficient condition in Corollary 2.
		\begin{align}
	\mathbb{E}\left[ \frac{1}{\mathcal{I}_i^2}\right]=\int\limits_{0}^{\infty}\int\limits_{0}^{\infty} 2 \pi \lambda_0 r t \exp \left( -\pi \lambda_0 r^2 -2 \pi \lambda_0  \int_{r}^{\infty}x \mathbb{E}_h \left[1-  e^{-h t L(x)} \right] dx  \right)drdt \notag \\
	\leq \int\limits_{0}^{\infty}\int\limits_{0}^{\infty} 2 \pi \lambda_0 r t \exp \left( -\pi \lambda_0 r^2 -2 \pi \lambda_0  \int_{r}^{\infty}x \mathbb{E}_h \left[1-  e^{-h t \tilde{L}(x)} \right] dx  \right)drdt. \notag 
		\end{align}
		Then the proof follows as in the proof of Corollary 5.
		\end{proof}
\end{corollary}  

So far we have proved that the {  average} ASE converges to a finite constant under fairly general assumptions. As discussed earlier, the definition of the {  average} ASE as in \eqref{Eq:ASE_Def1} assumes that we can work with any arbitrary small SINR. A more practical definition is the {  average} constrained ASE defined in \eqref{Eq:ASE_Def2} or perhaps the {  average} potential throughput defined in \eqref{Eq:ASE_Def3} in some scenarios. In the following, we leverage the analysis of the {  average} ASE to study the asymptotic behavior of the {  average} constrained ASE and the {  average} potential throughput.
\begin{theorem}
	If the path loss function satisfies the condition in Theorem 1 (or sufficiently Corollary 5) and $\theta_0 \geq \epsilon>0$, then,
	\begin{align}
		\lim_{\lambda \rightarrow \infty} \mathbb{E} \left[\tilde{\mathcal{E}}(\lambda,\theta_0)\right]=0,\notag
	\end{align}
	 where $\tilde{\mathcal{E}}(\lambda)$ is the {  average} constrained ASE defined in \eqref{Eq:ASE_Def2}.
	\begin{proof} 
		Let $\lambda=k \lambda_0$, where $k \in \mathbb{Z}_{+}$ and $\lambda_0 \in \mathbb{R}^{*}_{+}$. Then,
		\begin{align}
		\lim_{k \rightarrow \infty} \mathbb{E} \left[\tilde{\mathcal{E}}(k \lambda_0,\theta_0)\right]&=\lim_{k \rightarrow \infty} \lambda_0 k\mathbb{E} \left[ \log_2 (1+{\rm SINR})\mathbbm{1} \left\{{\rm SINR}\geq\theta_0 \right\} \right],\notag\\
		&=\mathbb{E} \left[\lim_{k \rightarrow \infty} \lambda_0 k \log_2 (1+{\rm SINR})\mathbbm{1} \left\{{\rm SINR}\geq\theta_0 \right\} \right],\label{eq:Th2_1}\\
			&=\mathbb{E} \left[ \frac{L_0 h_0}{2 \pi \ln(2) \gamma}\mathbbm{1} \left\{\lim_{k \rightarrow \infty}{\rm SINR}\geq\theta_0 \right\} \right]\label{eq:Th2_2}=0,
		\end{align}
		where \eqref{eq:Th2_1} follows since $\lambda_0 k\mathbb{E} \left[ \log_2 (1+{\rm SINR})\mathbbm{1} \left\{{\rm SINR}\geq\theta_0 \right\} \right]\leq \lambda_0 k\mathbb{E} \left[ \log_2 (1+{\rm SINR}) \right]$ which is uniformly integrable as proved in Theorem 1. Hence, we can push the limit inside the expectation \cite{Probability_Durrett10}. \eqref{eq:Th2_2} follows from \eqref{eq:Le0_6} and the fact that $\mathbbm{1} \left\{ x\geq \theta_0 \right\}$ is continuous at $x=0$ under the assumption that $\theta_0 \geq \epsilon >0$, and the last equality follows by using Lemma 1. Finally, since the result is independent of $\lambda_0$ we can conclude that $ \lim_{\lambda \rightarrow \infty} \mathbb{E} \left[\tilde{\mathcal{E}}(\lambda,\theta_0)\right]=0$.
	\end{proof}
	\end{theorem}

\begin{theorem}
	 If the path loss function satisfies the conditions in Theorem 1 (or sufficiently Corollary 5), then for all $\theta_0 \in \mathbb{R}_{+}$,
	 \begin{align}
	 \lim_{\lambda \rightarrow \infty}\mathbb{E} \left[\mathcal{R}(\lambda,\theta_0)\right]=0,
	 \end{align}
	 where $\mathcal{R}(\lambda)$ is the {  average} potential throughput defined in \eqref{Eq:ASE_Def3}.
	\begin{proof} 
		\begin{align}
	\lim_{\lambda \rightarrow \infty}\mathbb{E} \left[\mathcal{R}(\lambda,\theta_0)\right]&=\lim_{\lambda \rightarrow \infty} \lambda \log_2 (1+\theta_0) \mathbb{P} \left\{{\rm SINR}\geq\theta_0 \right\}, \notag\\
		&=
	\lim_{\lambda \rightarrow \infty} \lambda \log_2 (1+\theta_0) \mathbb{E} \left[ \mathbbm{1} \left\{{\rm SINR}\geq\theta_0 \right\}  \right], \label{eq:Th3_1}\\
		&\leq \lim_{\lambda \rightarrow \infty} \lambda\mathbb{E} \left[ \log_2 (1+{\rm SINR})  \mathbbm{1} \left\{{\rm SINR}\geq\theta_0 \right\}  \right], \notag\\
		&= \lim_{\lambda \rightarrow \infty} \mathbb{E} \left[\tilde{\mathcal{E}}(\lambda,\theta_0)\right]=0,\label{eq:Th3_2}
		\end{align}
		where \eqref{eq:Th3_2} follows from Theorem 2 assuming $\theta_0\geq \epsilon>0$. If $\theta_0=0$, then it is clear from \eqref{eq:Th3_1} that the limit is also zero.
	\end{proof}
\end{theorem}
Hence, the {  average} constrained ASE and the {  average} potential throughput have completely different behaviors than the {  average} ASE: we observe a densification plateau (it saturates to a constant) for the {  average} ASE and a densification collapse (it drops to zero) for the {  average} constrained ASE and the {  average} potential throughput. From a network throughput perspective, the ASE result means that although for very high densities the SINR approaches zero, the {  average} sum throughput of the users can still be higher than zero since there are many users in the network. In other words, the increase in the co-channel interference is fully balanced by the increase in the number of users using this channel. However, if we put a constraint on the minimum operational SINR, then many of these users will go into outage and the sum throughput of all users approaches zero. 
 
 From the users' application layer perspective, we can have the following two scenarios. If the users' share of the resources (bandwidth and time) is fixed, their rate gets very small when the density is high. However, the sum throughput is still non-negligible. Hence, if we have IoT devices or sensors with small data rate requirements, then the network can accommodate densification. On the other hand, if the rate demand is high (e.g. video streaming) then the network will not be able to satisfy their needs although their sum throughput is still relatively high. The second scenario is when the per user share of resources scales with the intensity, simply because more BSs means less load per BSs and bigger shares for each served user. In this case the {  average} ASE is an indication of the per user rate. Hence, the user can benefit from densification as long as it can work with (very) small SINRs, for example through low rate codes and/or spread spectrum techniques.
 
 Another insight from the previous theorems is that channel state information has to be available at the transmitter to be able to harvest gains from densifying the network. Theorem 3 shows that transmitting with a constant rate will drop the {  average} potential throughput to zero regardless of how small is the rate. The previous three theorems also explain the different conclusions in the literature regarding the asymptotic {  average} ASE: adopting different definitions with the same network model leads to totally different conclusions.

Note that the analysis in this work is based on the assumption that all BSs  always have users to serve. The other case where we have a finite user density and some BSs will be idle and silent can be handled as in \cite{The_Renzo16,Downlink_Yu13}. 
After establishing the analytical part of this work, we move to the next section where we apply the presented theorems to the commonly used bounded path loss models in the literature and we show that the framework is general enough to cover most the used models. 
\section{Case Studies}\label{Sec:Examples}

\begin{table}[]
	\centering
	\caption{Path loss functions.}
	\label{tb:path}
	\resizebox{\textwidth}{!}{%
	\begin{tabular}{|l|l|l|l|l|}
		\hline
		\rowcolor[HTML]{EFEFEE} 
		 Path loss function                                                            &   $\lim\limits_{\lambda \rightarrow \infty}\mathbb{E}\left[ \mathcal{E}(\lambda) \right]$      &   Parameters   &  $\tilde{L}(r)$       &$r_0$                                                          \\ \hline
	 $L_{1}(r)=A\min(c_0,r^{-\eta})$                                                            & $\frac{(\eta-2) c_0^{\frac{2}{\eta}}}{\eta \pi \ln(2)}$ & $A>0,c_0>0,\eta>2$ & $A r^{-\eta}$& $\max\left(1,c_0^{\frac{-1}{\eta}}\right)$\\ \hline
	 $L_{2}(r)=A (c_0+r)^{-\eta}$                       & $\frac{\eta^2-3\eta+2}{2 \pi \ln(2) c_0^2}$ &$A>0,c_0>0,\eta>2$& $A (c_0+r)^{-\eta}$    & 1                                                                 \\ \hline
	 $L_{3}(r)=A(c_0+r^{\eta})^{-1}$   &     $\frac{\eta \sin \left(\frac{2 \pi}{\eta}\right)}{2 \pi^2 \ln(2) c_0^{\frac{2}{\eta}}}$                                 & $A>0,c_0>0,\eta>2$     &      $A(c_0+r^{\eta})^{-1}$  &$\frac{c_0 (\eta-2)}{2}$                                                          \\ \hline
	 $L_{4}(r)=A(c_0^2+r^2)^{\frac{-\eta}{2}}$   &       $\frac{\eta-2}{2 \pi c_0^2  \ln(2)}$                         & $A>0,c_0>0,\eta>2$   &$A(c_0^2+r^2)^{\frac{-\eta}{2}}$   &1                                                                 \\ \hline
	 $L_{5}(r)=Ae^{-\alpha r^\beta} $ & $\frac{\beta \alpha^{\frac{2}{\beta}}}{2 \pi \ln(2) \Gamma\left(\frac{2}{\beta}\right)}$ & $A>0,2\geq\beta>0,\alpha>0$     & $Ae^{-\alpha r^\beta}$&1                                                                         \\ \hline
	\end{tabular}}
\end{table}

\subsection{Single-slope models}
 We start with the famous power-law path loss model where the power attenuation is captured by $L(r)=A r^{-\eta}$, where $A$ and $\eta$ are positive parameters \cite{Empirical_Hata80}. This function is clearly unbounded, since $L(r)\rightarrow \infty$ as $r \rightarrow 0$ which is not physically feasible. Hence it is not included in the class of functions we are considering. In fact, it was shown in \cite{A_Andrews11} that the SIR distribution is independent of the BS density, a property referred to as the SIR scale-invariance, for Rayleigh fading channels, and extended later for more general fading channels \cite{Unified_Trigui16,Modeling_AlAmmouri16}. Hence, by checking the definitions in \eqref{Eq:ASE_Def1}, \eqref{Eq:ASE_Def2}, and \eqref{Eq:ASE_Def3}, it is clear that all of these metrics will increase linearly with the BS density. Which is a totally different scaling law than the ones we have. However, by slightly modifying this path loss function, we can get path loss functions that have the desired properties in Definition \ref{Def:PLM}. Examples of such functions are $L_{1}(r)=A\min(c_0,r^{-\eta})$, $L_{2}(r)=A (c_0+r)^{-\eta}$, and $L_{3}(r)=A(c_0+r^{\eta})^{-1}$, where $c_0,\eta$, and $A$ are positive parameters. These functions are bounded and satisfy the desired properties mentioned in Section II. Moreover, these functions also satisfy the conditions in Corollary 5 as shown in Table I, where we provide the asymptotic {  average} ASE along with $\tilde{L}(\cdot)$ and $r_0$ needed to verify the conditions in Corollary 5. Note that both the {  average} constrained ASE and the {  average} potential throughput decay to zero for these functions according to theorems 2 and 3.

It was shown in \cite{Effect_Liu16} that the {  average} potential throughput -called the ASE in that paper- decays to zero assuming path loss functions of the form of $L_2(\cdot)$ and $L_3(\cdot)$ with $c_0=1$ and Rayleigh fading channels. Hence, it agrees with our results and we further proved that this observation holds for any fading distribution. Analysis of asymptotic {  average} ASE assuming path loss functions of the form of $L_1(\cdot)$, $L_2(\cdot)$, and $L_3(\cdot)$ were previously absent from the literature to the best of our knowledge.

Another interesting path loss function is $L_{4}(r)=A(c_0^2+r^2)^{\frac{-\eta}{2}}$ which captures the case where the elevation difference between the BSs and users is $c_0>0$ \cite{Downlink_Atzeni17}. This model has the desired properties in Definition \ref{Def:PLM} and also satisfies the conditions in Corollary 5 as shown in Table I, which also includes the asymptotic {  average} ASE under this model in a simple closed form that exactly shows how the {  average} ASE scales with the elevation difference $c_0$. {  The work in \cite{Stochastic_Filo17} considers this model for studying the performance of ultradense cellular networks with Rayleigh fading. The authors showed that the average ASE converges to a non-zero constant in the limit of $\lambda \rightarrow \infty$ and derived an upper bound of this constant for the special case of $\eta=4$ given by $\frac{2 \sqrt{3}}{9 \log(2) c_0^2}$. In this work, we derive the exact value of this constant for any $\eta$, and for any small-scale fading distribution that has a finite mean, which is given in Table 1. The limit reduces to $\frac{1}{\pi  \log(2) c_0^2}$ for the special case $\eta=4$. Moreover, it was argued in \cite{Stochastic_Filo17} that if the average constrained ASE is considered, then this constant depends on the chosen value of $\theta_0$. Our results indicate that the constrained ASE drops to zero asymptotically for any $\theta_0 >0$. Interestingly, the work in \cite{Stochastic_Filo17} shows that the same behavior for the ASE holds for the case of a regular deployment of BSs (hexagonal grid).}

\subsection{Multi-slope models}
Moving to a more general path loss model, the multi-slope path loss has been widely used in the literature \cite{Downlink_Zhang15,Wireless_Goldsmith05,Wireless_Rappaport01} and it is the baseline model in 3GPP standards \cite{3GPP2010,3GPP2017}. It can be represented as the following:
\begin{align}
L_{6}(r)=\sum_{i=1}^{n} A_i r^{-\eta_i} \mathbbm{1}\left\{r_{i-1}\leq r <r_{i} \right\},\notag
\end{align}
where $A_i,\eta_i$, $n$, and $r_i$ are positive parameters. The multi-slope path loss function captures the case where the path loss exponent $\eta$ is also a function of the distance. The number of slopes (regions) is $n$  with $\eta_i$  the effective path loss exponent in the $i^{\rm th}$ region that lies between $r_{i-1}$ and $r_i$. More details about this path loss function are provided in \cite{Downlink_Zhang15}. Without loss of generality, we set $r_0$ to zero and $r_n$ to $\infty$ to cover the whole positive real line and we assume that $n$ is at least $2$. Moreover, we set $\eta_1$ to zero to make it bounded and assume that $\eta_n>2$, otherwise the aggregate network interference will have an infinite average power \cite{Downlink_Zhang15}. Under these assumptions, this function has the desired properties in Definition 1, and by choosing $\tilde{L}(r)=A_n r^{-\eta_n}$ and $r_0=\max \left(1,r_{n-1} \right)$, the conditions in Corollary 5 are satisfied so that the asymptotic {  average} ASE is given by the following:
\begin{align} \label{eq:Multi_limit}
\lim\limits_{\lambda \rightarrow \infty}\mathbb{E}\left[ \mathcal{E}(\lambda) \right]= \frac{A_1}{2 \pi \ln(2)} \left(\frac{A_1 r_1^2}{2}+\frac{A_n r_{n-1}^{2-\eta_n}}{\eta_n}+\sum\limits_{i=2}^{n-1} A_i\left( \frac{r_i^{-\eta_i+2}}{2-\eta_i}+\frac{r_{i-1}^{-\eta_i+2}}{\eta_i-2}\right) \right)^{-1}.
\end{align}

However, both the {  average} potential throughput and the {  average} constrained ASE eventually drop to zero, which agrees with the results in \cite{Downlink_Zhang15}, where the authors showed that the {  average} potential throughput drops to zero in the case of $\eta_{1}=0$. Note that in the case of $n=2$ and $\eta_{1}=0$, the multi-slope model reduces to $L_1(\cdot)$ given in Table 1. {  The authors in \cite{Performance_Nguyen17} also consider this path loss model in studying the performance of ultradense cellular networks. Under the assumptions of a small-scale fading distribution that has finite mean and a bounded multi-slope path loss model, their results show that both the average ASE and the average potential throughput asymptotically scale sub-linearly with the BS density. Specifically, they showed that $\lim_{\lambda \rightarrow \infty} \frac{\mathbb{E}[\mathcal{E}(\lambda)]}{\lambda} \rightarrow 0$ and $\lim_{\lambda \rightarrow \infty} \frac{\mathbb{E}[\mathcal{R}(\lambda)]}{\lambda} \rightarrow 0$. Our results are in-line with the findings in \cite{Performance_Nguyen17}, but stronger since we proved that $\lim_{\lambda \rightarrow \infty} \mathbb{E}[\mathcal{E}(\lambda)] \rightarrow C$ and $\lim_{\lambda \rightarrow \infty} \mathbb{E}[\mathcal{R}(\lambda)] \rightarrow 0$, where $C$ is given by \eqref{eq:Multi_limit}. In addition, the work in \cite{Performance_Nguyen17} considers other cases where the small-scale fading has an infinite mean and/or the path loss in unbounded, but these results are out of the scope of this paper since we focused on practical assumptions where the small-scale fading has a finite mean and the path loss is bounded.}
%

\subsection{Stretched exponential model}
The limiting case of the multi-slope model with large number of slopes is captured by the stretched exponential path loss proposed in \cite{SINR_AlAmmouri17} to capture the power attenuation in urban and dense cellular networks where the signal loss is mainly due to obstructions. This path loss function takes the following form: $L_{5}(r)=Ae^{-\alpha r^\beta}$, where $A$, $\alpha$, and $\beta$ are positive parameters.
{  This model was originally proposed in \cite{A_Franceschetti04} as $Ae^{-\alpha r}r^{-\eta}$ and extended in \cite{Ray_Marano05} to $Ae^{-\alpha r^\beta}r^{-\eta}$, where $\eta$ is a positive parameter. However, it was shown in \cite{SINR_AlAmmouri17} that the term $r^{-\eta}$ is not effective in typical distance ranges in cellular networks (it is effective for very small distances $r<1$), hence it was removed to maintain the boundedness of the path loss function.}
  For more details about the intuition behind this model, refer to \cite{SINR_AlAmmouri17}. As shown in Table I, this model also satisfies the conditions in Corollary 5. It was proven in \cite{SINR_AlAmmouri17} that the {  average} ASE saturates to a constant and the {  average} potential throughput drops to zero for high BSs densities. Our results in this work confirm that both of these observations also hold for any general fading model. In addition, we prove that the {  average} constrained ASE also drops to zero.

\subsection{LoS/NLoS models}
 We have focused so far on path loss models that are deterministic with the distance. However, in the case of mmWave transmission, the power attenuation strongly depends on the link state (LoS or NLoS) \cite{Modeling_Andews16}. In fact, the path loss function adopted recently by the 3GPP standard \cite{3GPP2017} consider this scenario not only for mmWave, but for the whole frequency range from $0.5$ GHz to $100$ GHz. This path loss has three ingredients: multi-slope path loss, LoS/NLoS link states, and non-zero elevation difference between the BSs and the users. Mathematically it can be represented by the following:
\begin{align}
L(r,c_0)= P_{LoS}(r) L_{LoS}(r,c_0)+(1-P_{LoS}(r)) L_{NLoS}(r,c_0),\notag
\end{align}
with
\begin{align}
 L_{LoS}(r,c_0)&=\sum_{i=1}^{n} A_i (r^2+c_0^2)^{-\frac{\eta_i}{2}} \mathbbm{1}\left\{r_{i-1}\leq r <r_{i} \right\},\notag\\
  L_{NLoS}(r,c_0)&=\sum_{i=1}^{n} \bar{A}_i (r^2+c_0^2)^{-\frac{\bar{\eta}_i}{2}} \mathbbm{1}\left\{\bar{r}_{i-1}\leq r <\bar{r}_{i} \right\},\notag
\end{align}
 where, $P_{LoS}(r)$ is the LoS probability, $c_0$ is the elevation difference between the users and BSs, $A_i$ depends on the operating frequency, and the parameters associated with NLoS attenuation are distinguished by the bar above them. The LoS probability and all other parameters depend on the considered environment (e.g. RMa, UMa, UMi). However, regardless of the considered environment, we have the following characteristics of this path loss function:
 
 \begin{enumerate}
 	\item The LoS probability is decreasing with the distance. Hence, $L(0,c_0)=L_{LoS}(0,c_0)$. Moreover, $L_{LoS}(r,c_0)$ and $L_{NLoS}(r,c_0)$ are decreasing functions with the distance $r$. 
 	\item Due to the elevation difference between the users and the BSs, $L_{LoS}(0,c_0)=A_1 (c_0^2)^{\frac{-\eta_1}{2}}$ is bounded regardless of the value of $\eta_1$.
 	\item  For a given distance $r$, we always have $L_{LoS}(r,c_0) \geq  L_{NLoS}(r,c_0)$. Hence, $L(0,c_0) \geq L(r,c_0) \ \forall r \in \mathbb{R_{+}}$.
 	\item The values of $\eta_n$ and $\bar{\eta}_n$ are always larger than $2$ in all cases. In other words, the long distance path loss exponent is always larger than $2$, which is the path loss exponent for free space propagation. This ensures that $\gamma = \int\limits_{0}^{\infty} r L(r)dr$ is finite.
 \end{enumerate}

Based on the previous points, this path loss model satisfies the three properties in Definition \ref{Def:PLM}. Moreover, by choosing $\tilde{L}(r)=\bar{A_n} (r^2+c_0^2)^{\frac{-\bar{\eta}_n}{2}}$ along with $r_0=\max \left(1,\bar{r}_{n-1} \right)$, it is straightforward to see that the conditions in Corollary 5 are also satisfied. Hence, we can conclude that the {  average} ASE saturates to a constant, while the {  average} constrained ASE and the {  average} potential throughput drop to zero for high BS densities under this path loss model.

The models in \cite{Downlink_Atzeni17,Performance_Ding17,Performance_Ding16} consider similar path loss functions. In \cite{Downlink_Atzeni17}, it was shown that the {  average} potential throughput drops to zero eventually, and in \cite{Performance_Ding17} it was shown that the {  average} constrained ASE also drops to zero for high BSs densities. Our results agree with both of these works, and adds to them the fact that the {  average} ASE saturates to a constants and does not drop to zero, even for general small-scale fading models. {  On the other hand, the authors in \cite{Performance_Ding16} showed that the average constrained ASE might drop in the transition from the NLoS to LoS, but then  increases linearly with the BS density. However, the authors assume that the BSs and the users are on the same elevation with no restriction on $\eta_1$, which means the considered path loss function is no longer bounded. Our results in this work show that this optimistic scaling law is an artifact of the pole at origin on the considered path loss function, and if boundedness is enforced by setting $\eta_1$ to zero, then the average constrained ASE drops to zero.}

Overall, we have shown in this section that the conditions we have on the path loss functions are simple to check and yet general enough to capture the practical path loss models used in the literature and the standards. Moreover, the unified framework allows us to draw conclusions regarding the {  average} ASE, the {  average} constrained ASE, and the {  average} potential throughput at the same time by checking the same conditions. 

{  Note that in this work we have focused on the cellular architecture case, where the user receives its data only from its nearest BS and is hence protected from very strong interference. It would be interesting to study the effect of adopting such a bounded path loss on the scaling laws of other architectures like the ad hoc network. Most papers in the literature on ad hoc networks consider the power-law path loss model, which might lead to optimistic scaling laws as we have shown for the cellular case.}

\section{Conclusion}\label{Sec:Conc}
We have studied the {  average} asymptotic area spectral efficiency (ASE) in a downlink cellular network under ultradensification. Assuming a Poisson point process for the spatial distribution of the base stations, a general small-scale fading model, and a class of physically feasible path loss models, we provided a general framework to analyze the {  average} ASE. Our results show that the average ASE saturates to a constant when the base station (BS) density is large, while the {  average} constrained ASE and the {  average} potential throughput both collapse to zero.   These results show that there are fundamental physical limits to the throughput gains that can be harvested with densification, with the important caveats that interference is treated as noise and the active user density scales with the base station density.   The results are also all asymptotic with BS density and we do not study when these asymptotics kick in.  These caveats point to interesting future work for dense networks, such as characterizing at what density the {  average} ASE saturation or collapse manifest.  Considering the effects of advanced interference suppression techniques such as joint (over multiple BSs) transmission or decoding, or successive interference cancellation, would also be of interest.

\appendices
\section{Proof of Lemma 1}
Let $\lambda=k \lambda_0$, where $k \in \mathbb{Z}_{+}$ and $\lambda_0 \in \mathbb{R}^{*}_{+}$. We are interested in $\lim\limits_{k\rightarrow \infty} k\lambda_0{\rm SINR}(k\lambda_0)$ which is found by the following		
\begin{align}
\lim\limits_{k\rightarrow \infty} k\lambda_0 {\rm SINR}(k\lambda_0)&=\lim\limits_{k\rightarrow \infty}  k\lambda_0 \frac{h_0 L(r_0)}{ \sum\limits_{r_i\in \Phi \setminus B(0,r_0)}h_i L(r_i) +N_0}\notag\\
&=\lim\limits_{k\rightarrow \infty} k\lambda_0 \frac{h_0 L(r_0)}{ \sum\limits_{r_i\in \Phi }h_i L(r_i)-h_0 L(r_0) +N_0}\notag\\
&=\lim\limits_{k\rightarrow \infty} \lambda_0 \frac{h_0 L_0}{ \frac{1}{k}\sum\limits_{r_i\in \Phi }h_i L(r_i) -\frac{h_0 L_0}{k} +\frac{N_0}{k}}\label{eq:Le0_3}\\
&=\lim\limits_{k\rightarrow \infty} \lambda_0 \frac{h_0 L_0}{ \frac{1}{k}\sum\limits_{r_i\in \Phi }h_i L(r_i) }\label{eq:Le0_31}\\
&=\lim\limits_{k\rightarrow \infty}  \frac{ \lambda_0 h_0 L_0}{ \frac{1}{k}\sum\limits_{i=1}^{k} \sum\limits_{r_{i,j}\in \Psi_j }h_{i,j} L(r_{i,j}) }\label{eq:Le0_4}\\
&=  \frac{\lambda_0 h_0 L_0}{ \mathbb{E} \left[ \sum\limits_{r_{i,1}\in \Psi_1 }h_{i,1} L(r_{i,1}) \right]}\label{eq:Le0_5}\\
&=  \frac{\lambda_0 h_0 L_0}{2 \pi \lambda_0 \int\limits_{0}^{\infty} r L(r)dr  }=\frac{h_0 L_0}{2 \pi \gamma}\label{eq:Le0_6},
\end{align}
where, \eqref{eq:Le0_3}  follows because $L(r_0) \rightarrow L_0$ a.s. as $k \rightarrow \infty$, \eqref{eq:Le0_31} since as $k \rightarrow \infty$, $\frac{N_0}{k}\rightarrow 0$ and $\frac{h_0 L_0}{k}\rightarrow 0$ a.s, \eqref{eq:Le0_4}  by exploiting the superposition property of the PPP \cite{Stochastic_Baccelli10}, where $\Phi=\sum_{i=1}^{k} \Psi_i$ and $\Psi_i \ \forall i\in \mathbb{Z}_{+}^{*}$ are i.i.d. PPPs each with density $\lambda_0$, \eqref{eq:Le0_5} from the law of large numbers, and \eqref{eq:Le0_6}  from Campbell's  theorem\cite{Stochastic_Baccelli10}. Note that since the result is independent of $\lambda_0$, we can conclude that $\lim\limits_{\lambda \rightarrow \infty} \lambda{\rm SINR}(\lambda)=\frac{h_0 L_0}{2 \pi \gamma}$ which is finite a.s, hence $\lim\limits_{\lambda \rightarrow \infty}{\rm SINR}(\lambda)=0$ a.s, which completes the proof.

\section{Proof of Theorem 1}
	Since we proved the lower bound in Lemma 2, it is sufficient to prove that the asymptotic ASE is upper bounded by the constant given in \eqref{eq:Th1_1}. The proof is as follows:
\begin{align}
\lim\limits_{k \rightarrow \infty} \mathbb{E}\left[ \mathcal{E}(k\lambda_0) \right]&=\lim\limits_{k \rightarrow \infty} \mathbb{E}\left[ \lambda_0 k \log_2(1+{\rm SINR}(k\lambda_0)) \right]\notag\\
&\leq \lim\limits_{k \rightarrow \infty} \mathbb{E}\left[  \frac{\lambda_0}{\ln(2)} k {\rm SINR}(k\lambda_0) \right]\label{Eq:L2_2}\\
&=\lim\limits_{k \rightarrow \infty} \mathbb{E}\left[ \frac{\lambda_0}{\ln(2)} k \frac{h_0 L(r_0)}{ \sum\limits_{r_i\in \Phi \setminus B(0,r_0)}h_i L(r_i) +N_0}  \right]\notag\\
&\leq \lim\limits_{k \rightarrow \infty} \mathbb{E}\left[ \frac{\lambda_0 L_0}{\ln(2)} k \frac{h_0 }{ \sum\limits_{r_i\in \Phi \setminus B(0,r_0)}h_i L(r_i)}  \right]\label{Eq:Th1_5}\\
&= \frac{\lambda_0 L_0}{\ln(2)} \lim\limits_{k \rightarrow \infty} \mathbb{E}\left[  k \frac{1 }{ \sum\limits_{r_i\in \Phi \setminus B(0,r_0)}h_i L(r_i)}  \right]\label{Eq:Th1_6},
\end{align}
where, \eqref{Eq:L2_2} follows since $\ln(1+x)\leq x \ \forall x\in \mathbb{R_{+}}$, \eqref{Eq:Th1_5} follows because $L(r)<=L_0 \ \forall r\in \mathbb{R}_{+}$ and $N_0>0$, and \eqref{Eq:Th1_6} by taking the expectation with respect to $h_0$. Next, assuming that the RV inside the expectation in \eqref{Eq:Th1_6} is uniformly integrable (which we will prove next if $\mathcal{I}$ has a finite second negative moment), then we can push the limit inside the expectation\footnote{Precisely, if the RV inside the expectation in \eqref{Eq:Th1_6} is uniformly integrable, then a.s convergence implies convergence in the $L^{1}$ norm \cite[Theorem 5.5.2]{Probability_Durrett10}.}  \cite{Probability_Durrett10}.
\begin{align}
\frac{\lambda_0 L_0}{\ln(2)} \lim\limits_{k \rightarrow \infty} \mathbb{E}\left[   \frac{1 }{\frac{1}{k} \sum\limits_{r_i\in \Phi \setminus B(0,r_0)}h_i L(r_i)}  \right]=&\frac{\lambda_0 L_0}{\ln(2)}  \mathbb{E}\left[\lim\limits_{k \rightarrow \infty}   \frac{1 }{ \frac{1}{k}\sum\limits_{r_i\in \Phi \setminus B(0,r_0)}h_i L(r_i)}  \right]\notag\\
=&\frac{L_0}{\ln(2) 2 \pi \gamma}\label{Eq:Th1_8},
\end{align}
where \eqref{Eq:Th1_8} follows from the law of large numbers and Campbell's theorem similar to the proof in Lemma 1. Hence, the asymptotic average ASE is upper bounded by $\frac{L_0}{\ln(2) 2 \pi \gamma}$ and we already proved in Lemma 2 that it is lower bounded by this same constant. Moreover, since the results hold for all $\lambda_0 \geq \lambda_c \in \mathbb{R}_{+}$, then we can conclude that $\lim\limits_{\lambda \rightarrow \infty}\mathbb{E}\left[ \mathcal{E}(\lambda) \right]=\frac{L_0}{\ln(2) 2 \pi \gamma}$.

The final step is to prove that the RV inside the expectation in \eqref{Eq:Th1_6} is uniformly integrable if $\mathcal{I}$ has a finite second negative moment, which is proved as follows:
\begin{align}
\frac{1 }{ \frac{1}{k}\sum\limits_{r_i\in \Phi \setminus B(0,r_0)}h_i L(r_i)} &\leq \frac{1 }{\frac{1}{k}\sum\limits_{i=1}^{k} \sum\limits_{r_{i,j}\in \Psi_i \setminus B(0,r_{i,0})}h_{i,j} L(r_{i,j})} \label{Eq:L2_8}\\
&= \frac{1 }{\frac{1}{k}\sum\limits_{i=1}^{k} \mathcal{I}_i}\label{Eq:L2_9}\\
&\leq \frac{1}{k}\sum\limits_{i=1}^{k}  \frac{1 }{\mathcal{I}_i}\label{Eq:L2_10},
\end{align}
where, \eqref{Eq:L2_8} follows because in the LHS we exclude the closest node to the origin from the interference and in the RHS we exclude the closest point in each $\Psi_i$, where $\Psi_i \ \forall i$ are i.i.d. PPPs with intensity $\lambda_0$, so we remove the closest BS in $\Phi$ and additional ones, \eqref{Eq:L2_9} by defining $\mathcal{I}_i=\sum\limits_{r_{i,j}\in \Psi_i \setminus B(0,r_{i,0})}h_{i,j} L(r_{i,j})$, and \eqref{Eq:L2_10} follows from the convexity of $1/x$. Note that in \eqref{Eq:L2_10}, the random variables $\frac{1 }{\mathcal{I}_i}$ are i.i.d. and independent from the indexing $k$. Hence, if $\frac{1 }{\mathcal{I}_i}$ has a finite second moment, then the sequence of RVs (indexed by $k$) in \eqref{Eq:L2_10} has a finite second moment and is uniformly integrable \cite[Theorem 5.5.2]{Probability_Durrett10}, which implies that the sequence of RVs in the LHS of \eqref{Eq:L2_8} is also uniformly integrable and this concludes the proof.
\section{Proof of Corollary 1}

The second negative moment can be written as:
\begin{align}
\mathbb{E}\left[ \frac{1}{\mathcal{I}_i^2}\right]&=\int\limits_{0}^{\infty} t \mathcal{M}_{\mathcal{I}}(-t)dt,\notag
\end{align}
where $\mathcal{M}_{\mathcal{I}}(\cdot)$ is the moment generating function of the interference $\mathcal{I}$. Using classical stochastic geometry analysis \cite{A_Andrews11}, it can be represented as the following
\begin{align}
\mathcal{M}_{\mathcal{I}}(t)&=\mathbb{E} \left[e^{t I} \right]=\mathbb{E} \left[e^{t \sum_{r_i \in \Psi\setminus B(0,r_0)} L(r_i) h_i} \right]\notag\\
&=\int_{0}^{\infty}\mathbb{E} \left[ \prod_{r_i\in \Psi\setminus B(0,r_0)} e^{t L(r_i) h_i} |r_0\right] 2 \pi \lambda_0 e^{- \pi \lambda_0 r_0^2} dr_0\notag\\
&=\int_{0}^{\infty} \exp \left( -2 \pi \lambda_0  \int_{r}^{\infty}x \left(1- \mathbb{E}_h \left[ e^{h t L(x)} \right] \right)dx \right) 2 \pi \lambda_0 e^{- \pi \lambda_0 r_0^2} dr_0, \label{eq:Le2_11}
\end{align}
where \eqref{eq:Le2_11} follows from the probability generating functional (PGFL) of a PPP \cite{Stochastic_Baccelli10_2}. Hence, the second negative moment of $\mathcal{I}$ is given by
\begin{align}
\mathbb{E}\left[ \frac{1}{\mathcal{I}_i^2}\right]&=\int\limits_{0}^{\infty}\int\limits_{0}^{\infty} 2 \pi \lambda_0 r t \exp \left( -\pi \lambda_0 r^2 -2 \pi \lambda_0  \int_{r}^{\infty}x \left(1- \mathbb{E}_h \left[ e^{-h t L(x)} \right] \right)dx  \right)drdt\label{Eq:Le2_1}.
\end{align}

From \eqref{Eq:Le2_1}, we get the first sufficient condition. The second condition can be found by the following:		
\begin{align}
&\mathbb{E}\left[ \frac{1}{\mathcal{I}_i^2}\right]=	\mathbb{E}\left[ \left(\sum\limits_{r_i \in \Psi \setminus B(0,r_{0})} L(r_i) h_i \right)^{-2}\right]\leq 	\mathbb{E}\left[\left( \max\limits_{r_i \in \Psi\setminus B(0,r_{0})}  L(r_i) h_i \right)^{-2}\right]\notag\\
&=\int\limits_{0}^{\infty} \mathbb{P} \left(\left( \max\limits_{r_i \in \Psi \setminus B(0,r_{0})}  L(r_i) h_i \right)^{-2}>t\right)dt\notag\\
&=\int\limits_{0}^{\infty} \mathbb{P} \left( \max\limits_{r_i \in \Psi \setminus B(0,r_{0})}  L(r_i) h_i <\frac{1}{\sqrt{t}}\right)dt\notag\\
&=\int\limits_{0}^{\infty} \mathbb{E} \left[\mathbbm{1}\left\{\max\limits_{r_i \in \Psi \setminus B(0,r_{0})}  L(r_i) h_i <\frac{1}{\sqrt{t}}\right\}\right]dt\notag\\
&=\int\limits_{0}^{\infty} \mathbb{E} \left[ \prod\limits_{r_i \in \Psi \setminus B(0,r_{0})} \mathbbm{1}\left\{  L(r_i) h_i <\frac{1}{\sqrt{t}}\right\}\right]dt=\int\limits_{0}^{\infty} \mathbb{E} \left[ \prod\limits_{r_i \in \Psi \setminus B(0,r_{0})} e^{\log \left(\mathbbm{1}\left\{  L(r_i) h_i <\frac{1}{\sqrt{t}}\right\}\right)}\right]dt\notag\\
&=\int\limits_{0}^{\infty}\int\limits_{0}^{\infty}2 \pi \lambda_0 r \exp \left(- \pi \lambda_0 r^2- 2 \pi \lambda_0 \int\limits_{r}^{\infty}x \mathbb{E}_h \left[1-\mathbbm{1} \left\{L(x) h <\frac{1}{\sqrt{t}} \right\} \right]dx \right) drdt\label{Eq:Co1_1}\\
&=\int\limits_{0}^{\infty} \int\limits_{0}^{\infty}2 \pi \lambda_0 r \exp \left(- \pi \lambda_0 r^2- 2 \pi \lambda_0 \int\limits_{r}^{\infty}x  \left[1- \mathbb{P}_h\left(L(x) h <\frac{1}{\sqrt{t}} \right) \right]dx \right) drdt\label{Eq:Co1_2},
\end{align}
where \eqref{Eq:Co1_1} follows from the PGFL of a PPP \cite{Stochastic_Baccelli10}.
\section{Proof of Corollary 2}
For the Rayleigh fading case, $\mathbb{E}_h \left[1-  e^{-h t L(x)} \right]=\frac{t L(x)}{1+tL(x)}$. By substituting this in the first condition in the last corollary we get
\begin{align}
&\int\limits_{0}^{\infty}\int\limits_{0}^{\infty} 2 \pi \lambda_0 r t \exp \left( -\pi \lambda_0 r^2 -2 \pi \lambda_0  t \int_{r}^{\infty}\frac{x L(x)}{1+tL(x) }dx \right)drdt=T_1+T_2,\notag\end{align}
with
\begin{align}
&T_1=\int\limits_{0}^{\infty}\int\limits_{0}^{r_0} 2 \pi \lambda_0 r t \exp \left( -\pi \lambda_0 r^2 -2 \pi \lambda_0  t \int_{r}^{\infty}\frac{x L(x)}{1+tL(x) }dx \right)drdt\notag\\
&T_2=\int\limits_{0}^{\infty}\int\limits_{r_0}^{\infty} 2 \pi \lambda_0 r t \exp \left( -\pi \lambda_0 r^2 -2 \pi \lambda_0  t \int_{r}^{\infty}\frac{x L(x)}{1+tL(x) }dx \right)drdt.\label{eq:Cor2T2}
\end{align}
Then if $T_1$ and $T_2$ are finite, we prove the corollary. Let $y_0 = L(r_0)$, then

\begin{align}
T_1&=\int\limits_{0}^{\infty}\int\limits_{0}^{r_0} 2 \pi \lambda_0 r t \exp \left( -\pi \lambda_0 r^2 -2 \pi \lambda_0  t \int_{r}^{\infty}\frac{x L(x)}{1+tL(x) }dx \right)drdt\notag\\
&\leq\int\limits_{0}^{\infty}\int\limits_{0}^{r_0} 2 \pi \lambda_0 r t \exp \left( -\pi \lambda_0 r^2 -2 \pi \lambda_0  t \int_{r_0}^{\infty}\frac{x L(x)}{1+tL(x) }dx \right)drdt\notag\\
&=\int\limits_{0}^{\infty}\int\limits_{0}^{r_0} 2 \pi \lambda_0 r t \exp \left( -\pi \lambda_0 r^2 -2 \pi \lambda_0  t \int_{0}^{y_0}\frac{L^{-1}(y)y}{-L^{'}(L^{-1}(y))} \frac{1}{1+ty }dy \right)\label{eq:Cor2_1}drdt\\
&\leq\int\limits_{0}^{\infty}\int\limits_{0}^{r_0} 2 \pi \lambda_0 r t \exp \left( -\pi \lambda_0 r^2 -2 \pi \lambda_0  t \int_{0}^{y_0} \frac{\zeta}{1+ty }dy \right)drdt\label{eq:Cor2_2}\\
&=\int\limits_{0}^{\infty}\int\limits_{0}^{r_0} 2 \pi \lambda_0 r t \exp \left( -\pi \lambda_0 r^2 -2 \pi \lambda_0 \zeta \log(1+t y_0) \right)drdt\notag\\
&=\int\limits_{0}^{\infty}\int\limits_{0}^{r_0} 2 \pi \lambda_0 r t \exp \left( -\pi \lambda_0 r^2\right) \frac{1}{\left( 1+t y_0 \right)^{2 \pi \lambda_0 \zeta}} drdt\notag\\
&=(1-e^{-\pi \lambda_0 r_0^2})\int\limits_{0}^{\infty} \frac{t}{\left( 1+t y_0 \right)^{2 \pi \lambda_0 \zeta}} dt\label{eq:Cor2_3},\\
&=\frac{1-e^{-\pi \lambda_0 r_0^2}}{y_0^2 (2 \pi \lambda_0 \zeta-1)(2 \pi \lambda_0 \zeta-2)}\label{eq:Cor2_31},
\end{align}
where, \eqref{eq:Cor2_1} follows by the substitution $y=L(r)$, \eqref{eq:Cor2_2} since $\frac{rL(r)}{-L^{'}(r)} \geq \zeta ,\  \forall r \in [r_0,\infty)$ which means that  $\frac{L^{-1}(y)y}{-L^{'}(L^{-1}(y))} \geq \zeta ,\  \forall y \in (0,y_0=L(r_0)]$. Finally, \eqref{eq:Cor2_31} holds as long as $\lambda_0>\frac{1}{\pi \zeta}$. Hence $T_1$ is finite. Moving to $T_2$ defined in \eqref{eq:Cor2T2},


\begin{align}
T_2&=\int\limits_{0}^{\infty}\int\limits_{r_0}^{\infty} 2 \pi \lambda_0 r t \exp \left( -\pi \lambda_0 r^2 -2 \pi \lambda_0  t \int_{r}^{\infty}\frac{x L(x)}{1+tL(x) }dx \right)drdt\notag\\
&=\int\limits_{0}^{\infty}\int\limits_{r_0}^{\infty} 2 \pi \lambda_0 r t \exp \left( -\pi \lambda_0 r^2 -2 \pi \lambda_0  t \int_{0}^{L(r)}\frac{L^{-1}(y)y}{-L^{'}(L^{-1}(y))} \frac{1}{1+ty }dy \right)drdt\notag\\
&\leq \int\limits_{0}^{\infty}\int\limits_{r_0}^{\infty} 2 \pi \lambda_0 r t \exp \left( -\pi \lambda_0 r^2 -2 \pi \lambda_0  t \int_{0}^{L(r)}\frac{\zeta}{1+ty }dy \right)drdt \label{eq:Cor2_4}\\
&=\int\limits_{0}^{\infty}\int\limits_{r_0}^{\infty} 2 \pi \lambda_0 r t \exp \left( -\pi \lambda_0 r^2\right) \frac{1}{\left( 1+t L(r) \right)^{2 \pi \lambda_0 \zeta}} drdt\notag\\
&=\int\limits_{r_0}^{\infty} 2 \pi \lambda_0 r  \exp \left( -\pi \lambda_0 r^2\right)\int\limits_{0}^{\infty} \frac{t}{\left( 1+t L(r) \right)^{2 \pi \lambda_0 \zeta}} dtdr\notag\\
&=\int\limits_{r_0}^{\infty} 2 \pi \lambda_0 r  \exp \left( -\pi \lambda_0 r^2\right)\frac{1}{b L(r)^2}dr\notag\\
&=\frac{2 \pi \lambda_0 }{b}\int\limits_{r_0}^{\infty} \frac{r}{ L(r)^2}  \exp \left( -\pi \lambda_0 r^2\right)dr,\label{eq:Cor2_5}
\end{align}
where \eqref{eq:Cor2_4} follows because $\frac{L^{-1}(y)y}{-L^{'}(L^{-1}(y))} \geq \zeta ,\  \forall y \in (0,y_0=L(r_0)]$ and  \eqref{eq:Cor2_5} holds as long as $\lambda_0 \geq \frac{1}{\pi \zeta}$ with $b=(2 \pi \lambda_0 \zeta-2)(2 \pi \lambda_0 \zeta-1)$. The last integral is finite by assumption. Hence, $T_2$ is also finite which proves the corollary.

\section{Proof of Corollary 4}
Define $\tilde{\mathcal{I}}=\mathcal{I}/2=\sum_{r_i \in \Psi\setminus B(0,r_0)} L(r_i) \tilde{h}_i$, where $\tilde{h}_i=h_i/2$, Clearly, if $\tilde{\mathcal{I}}$ has a finite second negative moment, then $\mathcal{I}$ will have a finite second negative moment. 
\begin{align}
\mathbb{E}\left[ \frac{1}{\tilde{\mathcal{I}}^2}\right]&=\int\limits_{0}^{\infty}\int\limits_{0}^{\infty} 2 \pi \lambda_0 r t \exp \left( -\pi \lambda_0 r^2 -2 \pi \lambda_0  \int_{r}^{\infty}x \left(\mathbb{E}_{\tilde{h}} \left[1-  e^{-\tilde{h} t L(x)} \right] \right)dx  \right)drdt\notag\\
\leq&\int\limits_{0}^{\infty}\int\limits_{0}^{\infty} 2 \pi \lambda_0 r t \exp \left( -\pi \lambda_0 r^2 -2 \pi \lambda_0 t  \int_{r}^{\infty}\mathbb{E}_{\tilde{h}} \left[x  \frac{\tilde{h}  L(x)}{1+\tilde{h} t L(x)} \right] dx \right)drdt\label{Eq:Le5_2},
\end{align}
where \eqref{Eq:Le5_2} follows since $1-e^{-t}\geq \frac{t}{1+t}, \ \forall t\geq0$. Moreover, 
\begin{align}
\mathbb{E}_{\tilde{h}} \left[  \frac{x\tilde{h}  L(x)}{1+\tilde{h} t L(x)} \right]&=\int\limits_{0}^{\infty}   \frac{x\tilde{h}  L(x)}{1+\tilde{h} t L(x)} f_{\tilde{H}}(\tilde{h})d\tilde{h} \notag \\
&\geq\int\limits_{0}^{1}   \frac{x\tilde{h}  L(x)}{1+\tilde{h} t L(x)} f_{\tilde{H}}(\tilde{h})d\tilde{h}\label{Eq:Le5_4}\\
&\geq\int\limits_{0}^{1}   \frac{x\tilde{h}  L(x)}{1+ t L(x)} f_{\tilde{H}}(\tilde{h})d\tilde{h}=  \frac{x  L(x)}{1+ t L(x)} \mathbb{P} \left\{\tilde{h}\leq 1 \right\}\notag\\
&\geq \frac{1}{2}  \frac{ x L(x)}{1+ t L(x)}\label{Eq:Le5_6},
\end{align}
where $f_{\tilde{H}}(\cdot)$ is the probability density function (PDF) of $\tilde{h}$. \eqref{Eq:Le5_4} follows since the integrand is positive and \eqref{Eq:Le5_6} since by Markov's inequality $\mathbb{P} \left\{\tilde{h}\leq 1 \right\}\geq 1-\mathbb{E}[\tilde{h}]=\frac{1}{2}$. Hence,
\begin{align}
\mathbb{E}\left[ \frac{1}{\tilde{\mathcal{I}}^2}\right]\leq\int\limits_{0}^{\infty}\int\limits_{0}^{\infty} 2 \pi \lambda_0 r t \exp \left( -\pi \lambda_0 r^2 - \pi \lambda_0 t  \int_{r}^{\infty}\frac{ x L(x)}{1+ t L(x)} dx \right)\notag drdt .
\end{align}

Then the proof follows as in the proof of Corollary 2.
\bibliographystyle{IEEEtran}
\bibliography{AhmadRef,Andrews}
\vfill
\end{document}